\documentclass[aps,amsfonts,prd,twocolumn]{revtex4}
\input{psfig.sty}
\newcommand{\ar}{\arrowvert}

\newcommand{\ov}{\overline}
\newcommand{\cd}{\! \cdot \!}
\newcommand{\be}{\begin{equation}}
\newcommand{\ee}{\end{equation}}
\newcommand{\ba}{\begin{eqnarray}}
\newcommand{\ea}{\end{eqnarray}}

\begin{document}
\title{The Viscosity of Meson Matter. \\}

\author{Antonio Dobado} \email{dobado@fis.ucm.es}
\author{Felipe J. Llanes-Estrada \footnote{On leave at 
the University of T\"ubingen, Inst. f\"ur Theoretische Physik, Auf der 
Morgenstelle 14,  D-72076  T\"ubingen, Germany.}}
\email{fllanes@fis.ucm.es}
\affiliation{Departamento de F\'{\i}sica Te\'orica I,  Universidad
Complutense, 28040 Madrid, Spain}

\date{\today}

\begin{abstract}
We report a calculation of the shear viscosity in a relativistic
multicomponent
meson gas as a function of temperature and chemical
potentials. We approximately solve the Uehling-Uhlenbeck transport
equation  of kinetic theory, appropriate for a boson gas, with
relativistic kinematics. Since at low
temperatures the gas can be taken as mostly composed of 
pions, with a fraction of kaons and etas, we explore the region where
binary elastic collisions with at least one pion are the dominant
scattering processes. Our input meson scattering phase shifts are
fits to the experimental data
obtained from chiral  perturbation theory and the Inverse Amplitude 
Method. Our results take the correct non-relativistic limit (viscosity
proportional to the square root of the temperature), show 
a viscosity of order the cubed of the pion mass up to temperatures 
somewhat below that mass, and then a large increase due to kaons and etas.
Our approximation may break down at even higher temperatures, where 
the viscosity follows a temperature power law with an exponent near 3.
\end{abstract}

\maketitle

\section{Introduction}
The conceptual framework of this paper is a hadronic medium 
at zero baryon number, and dilute enough that it can be considered
as a meson gas, not too different from a perfect fluid. 
We have good  reasons to believe this is a satisfactory approximation to 
the state  of matter in the debris following a relativistic heavy
ion collision (RHIC) in the laboratory, and it may also be relevant in
future astrophysical considerations, where the relatively simple
extension of this work to include nucleons is definitely of importance. 
The observable particle multiplicities, correlations, angular
distributions, etc., measured in RHIC, are customarily fitted to a few 
parameters
within hydrodynamical models for the exploding gas \cite{kolb}.
These include typically a set of initial conditions (phase transition
temperature after a putative quark and gluon plasma, QGP, reaches the
confined stage, initial energy density and state equation, etc.).
The hydrodynamical evolution is supposed to control the evolution
of the gas through a chemical freeze-out temperature, after which
the particle composition is fixed and chemical potentials become
necessary, down to a thermal freeze-out temperature, where
the hadrons suddenly abandon their equilibrium state within the
fluid and travel unscattered to the detector. This sudden transition
from perfect local equilibrium to straight particle streaming 
is known as the Cooper-Frye prescription.
Whether a ``geometric'' (related to the finite size of the expanding 
almond-shaped  region) or a ``dynamical'' (related to the local 
expansion rate) freeze-out, the scattering between individual particles 
governs the mean free path which would make sensible a hydrodynamical 
approach.

Mesons at low transverse momentum seem to behave hydrodynamically
in the sense that the observed momentum distributions can be
fit with hydrodynamic models up to a decoupling temperature
\cite{heinz}. Still the fits are worse for high-$p_T$ and 
observables within HBT (Hanbury, Brown, Twiss) interferometry and the 
elliptic flow coefficient seem to start requiring viscous corrections,
at least within the QGP \cite{teaney}. A brief, pedagogical
account of some of these issues can be found in \cite{pisarski}.

We expect a better approximation to be borne out by considering
corrections to the equations of a perfect fluid, allowing for 
a smoother transition to a free particle stream instead of 
a sharp cutoff at a given temperature. The ideal hydrodynamical
description requires the time scales of microscopic processes
to be much faster than the corresponding macroscopic fluid scale.
This is not necessarily true for a dilute pion gas at moderate
temperatures. For example, in \cite{goity}, the pion mean free
path is found to be as long as 4-5 fermi before a
thermal freeze-out at temperatures of order 90-110 MeV (conservatively 
low), with the chemical  freeze-out temperatures in the range 90-140 MeV.

Thus if a gradient in the concentration of a conserved quantity is present 
in the medium, it can be smoothed out by pions flying randomly without
colliding too often. This induces a need for transport coefficients in 
the hydrodynamical equations, in particular viscosity when the
conserved quantity is momentum. The larger the interaction, the 
shorter the mean free path, and therefore the smaller the viscosity (for
a dilute gas) 
as a consequence it is interesting to also document the sensitivity to the 
choice of parameterization of the pion interaction given in terms of 
diverse scattering phase shift sets.

To approach a microscopic calculation of the transport 
coefficients one can start from rigorous quantum field theory 
\cite{aarts} with all the generality of Green's functions
coupled to each other through Schwinger-Dyson equations.
In the problem at hand collisions are mostly elastic, not requiring a 
coupling between states with different number of particles, and as the
mean free paths are relatively large, we can employ Boltzmann's molecular
chaos hypothesis, or decorrelation between successive collisions. 
It is more reasonable in this case to employ the statistical formulation 
in terms of distribution functions $f$, defined below in 
section \ref{variousdefs}. These also 
satisfy  coupled equations with the multiparticle distribution functions
(BBKGY hierarchy of equations, after Bogoliubov, Born, Kirkwood, Green 
and Yuan) which the Boltzmann hypothesis truncates.
Quantum and relativistic effects need to be taken into account
for a meson gas. 
We thus set ourselves the task of solving the Uehling-Uhlenbeck 
equation \cite{uehling} with relativistic kinematics. The non-relativistic 
formulation  has been published in \cite{silvia}. The relativistic pion 
gas has been treated before by \cite{davesne} at the quantum level and by
\cite{prakash2} at the classical (Boltzmann) level as the viscosity is
concerned. 
We made a brief preliminary comment about this system in \cite{bienal}. We 
here report the full calculation and include the obvious extension to a 
gas  composed of pions and particles with strange quarks (kaons and etas). 
For this purpose we include pions, kaons and etas, which populate
the low temperature meson gas. The other possible states,
the  rho, sigma and kappa mesons, appear through resonances in meson 
scattering in our calculation. Consistent with a sort of virial
expansion \cite{gerber} we ignore collisions between kaons and
etas, and only include those between them and the pions. This is
valid at moderate temperatures because their density is small (they 
conform a very dilute gas up to temperatures of order 150 MeV or
more \cite{gerber}).

\section{Notation and Kinematics.}\label{variousdefs}
We gather here our conventions and notation for the rest of the paper,
many borrowed from \cite{landau}.
Since we will be considering meson matter at low temperatures,
composed mainly by pions, kaons and eta mesons, latin indices from
the beginning of the alphabet
$a$, $b$,... denote the particle type and  will take the values 
$a=\pi,K,\eta$ in the
formulae below. Notice that tensor quantities such as $\tau_{ij}$ also
carry latin indices in the range $1,2,3$ for Cartesian coordinate labels,
which will be denoted with the letters $i,j,k,...$. For these tensors,
a tilde $\tilde{V}_{ij}$ means  their traceless part, 
as is common use in textbooks. Greek indices are reserved for Minkowski 
space and run over the values $0,1,2,3$. The metric $g_{\alpha \beta}$ 
will be taken as ${\rm diag}(+---)$ ($\eta$ is reserved for the 
viscosity).

In the isospin limit which we employ the degeneracy of each species
of particle in the gas is
\be
g_\pi=3 \ \ g_K=4 \ \ g_\eta=1 \ .
\ee
We will work in the approximation (related for example to the virial 
expansion in \cite{gerber}) in which the particle densities $dN_a/dV$ for
each species satisfies
\be
n_\pi>>n_K\ , \ n_\eta \ .
\ee
Therefore we consider binary collisions between pion pairs or between a 
pion and either a kaon or an eta meson. This amounts to neglecting
terms of order $n_K^2 \ , n_\eta^2$.
For the elastic collisions we consider, $\pi \pi \longrightarrow \pi 
\pi$, $\pi K\longrightarrow \pi K$, $\pi \eta \longrightarrow \pi \eta$
the particle numbers $N_\pi$, $N_K$, $N_\eta$ are separately
conserved. This is a good approximation in the hadron gas following
a relativistic collision after chemical freeze-out. This conservation
forces the introduction of chemical potentials collectively denoted
$\mu_a$ and fugacities $z_a=e^{\beta(\mu_a-m_a)}$. We introduce
no baryon chemical potential as we work at zero baryon number.
The corresponding elastic cross-sections for meson scattering will be 
denoted by
$$
d\sigma_{ab}=d\sigma(ab\longrightarrow ab)
$$
with the following momentum assignments for the initial and final
states:
\begin{center}
\be
\begin{picture}(100,100)(0,0)
\put(0,100){\line(1,-1){50}}
\put(25,75){\vector(1,-1){10}}
\put(5,75){$a,\vec{p}$}
\put(0,0){\line(1,1){50}}
\put(25,25){\vector(1,1){10}}
\put(5,25){$b,\vec{p_1}$}
\put(50,50){\line(1,1){50}}
\put(60,60){\vector(1,1){10}}
\put(85,75){$a,\vec{p} \ '$}
\put(50,50){\line(1,-1){50}}
\put(60,40){\vector(1,-1){10}}
\put(85,25){$b,\vec{p_1} \ '$}
\end{picture}
\ .
\ee
\end{center}
Further, $\rho_{ab}$ is the 2-body phase
space 
$$
\rho_{ab}=\frac{1}{s}\sqrt{s^2+m_a^4+m_b^4-2m_a^2m_b^2-2sm_a^2-2sm_b^2}
$$
in terms of the Mandelstam variable $s=(p+p_1)^2$. The variable
$t=(p-p')^2$ is related to the scattering angle in the center of 
momentum frame: 
\be \label{scangle}
\cos \theta_{\rm CM} = 1+ \frac{2t}{s\rho_{ab}^2} 
\ee
which is used below in formulae (\ref{pipiamplitude}) and following.
The total momentum is $\vec{P}=\vec{p}+\vec{p}_1$ and the total
energy $E$.
Since we will choose a frame where the fluid is locally at rest (see
below) the collision needs to be taken in an arbitrary frame respect to 
which all angles and momenta will henceforth be referred as the collision 
is concerned. This also leads to an additional difficulty due to a
double valuedness in eq. (\ref{dossols}) below which is not present in
the center of mass frame usually employed in meson-meson scattering. 
The distribution functions in phase space will be denoted by
$f_a$, $f_b$, $f_a'$, $f_b'$, and they are shorthands for
$$
f_a=f_a(\vec{x},\vec{p};t) \ .
$$
The normalization constants for these functions ($\hbar=1$) are
$$
\xi_a=\frac{(2\pi)^3}{g_a} \ .
$$

\section{Theoretical Background.}
\subsection{Hydrodynamics.}
We start by writing all magnitudes for only one particle species. The
generalization to the three types of meson considered is obviously
additive in this subsection and will be understood.
The energy-stress tensor for an ideal fluid with local velocity field
$U^\alpha(x)$ is
\be
T_{\alpha \beta}= -P g_{\alpha \beta} + \omega U_\alpha U_\beta
\ee
with the enthalpy per unit of proper volume (comoving volume at
velocity $U$ respect to the fixed Lagrangian reference frame) being
the sum of pressure and energy density (also per proper volume):
$$
\omega=P+\rho \ .
$$
Also conserved in the fluid's evolution is the vector field
associated to the particle density flow:
\be
n_\alpha=nU_\alpha \ ,
\ee
where again the particle density is taken per unit of proper volume.
The velocity field $U$ as seen from the fixed reference frame
can be interpreted in terms of the three-velocity $\vec{V}$ by 
introducing the $\gamma$ factor 
$\gamma=(\sqrt{1-\vec{V}\cd\vec{V}})^{-1}$ as
$$
U=\gamma (1,\vec{V}) \ .
$$
All quantities shall by default be refered to the comoving or Eulerian 
fluid frame, where $\vec{V}=0$.
The ideal fluid continuity equation is
\be \label{continuity}
\partial_t(n\gamma) + \vec{\nabla}(n\gamma \vec{V})=0 \ .
\ee

If now we assume the fluid to be slightly out of equilibrium 
microscopically, that is, a small departure from the ideal fluid,
the conserved quantities need to be modified to allow for the various
transport phenomena: in a gas particles can move from one element of the 
fluid to another at a microscopic level, carrying their charge, particle
number, energy, momentum, etc. and maximizing entropy, they smooth
out the gradients of these quantities in the fluid.
The macroscopic description of these phenomena is achieved  by 
adding transport terms to the conserved currents and tensors. In 
particular, for the stress-energy tensor we add a $\tau$:
\be
T_{\alpha \beta} = -Pg_{\alpha \beta} + \omega U_\alpha U_\beta
+\tau_{\alpha \beta}
\ee
with
$$
\tau_{\alpha \beta}U^\beta=0
$$
since these transport phenomena act across fluid elements, and out
of the world-line of one of them.
This implies that in the Eulerian frame, $\tau_{i0}=
\tau_{0i}=\tau_{00}=0$.

To first  order in the velocity gradients (that is, for small spatial 
variations of the velocity field) $\tau$ is 
\ba \label{3Dstresstensor}
\tau_{ij}= -2\eta\tilde{V}_{ij} +\ {\rm volume \ term}\\ \nonumber
\tilde{V}_{ij} =\frac{1}{2} \left( \partial_i V_j+\partial_j V_i
\right) -\frac{1}{3} \partial_k V^k \delta_{ij} \\ \nonumber
\sum_i \tilde{V}_{ii}=0 \ .
\ea
The volume viscosity $\xi$ is usually much smaller than the shear 
viscosity $\eta$, this being the reason why we concentrate on the later. 
This has been shown for a pure pion gas by Davesne in \cite{davesne}. We
expect this result to hold in the multicomponent extension of 
the theory and accept it henceforth.
For completeness we give the expression for $\tau$ in a general frame
of reference:
\ba \nonumber
\tau_{\alpha \beta}=-\eta_s \left[ \partial_\beta U_\alpha + 
\partial_\alpha U_\beta- U_\beta U^\gamma \partial_\gamma U_\alpha \right.
\\ \nonumber \left.
- U_\alpha  U^\gamma \partial_\gamma U_\beta +\frac{2}{3}\partial_\gamma
U^\gamma(g_{\alpha \beta} -U_\alpha U_\beta)\right]
\ea
At the hydrodynamic level of a fluid's description, $\eta$ is
an empirical parameter. Its value needs to be measured (fitted)
for the different experimental conditions of the considered fluid.
\subsection{Thermodynamics.}
We collect in this subsection the relevant thermodynamical
properties leading to the equation of state of an ideal multi-component
Bose gas, Since we are interested in the leading viscosity effects,
we take the thermodynamical quantities to be unaffected by the
interactions. These could be corrected if interest be found in it
with the method of the virial expansion and the physical phase shifts
described in \cite{prakash2}
entailing the chemical potential $\mu_a$ for each species,
the particle mass $m_a$, the temperature $T$ and inverse
temperature $\beta$, fugacity $z_a$, and pressure P.

\begin{widetext}
\begin{figure}[h]
\psfig{figure=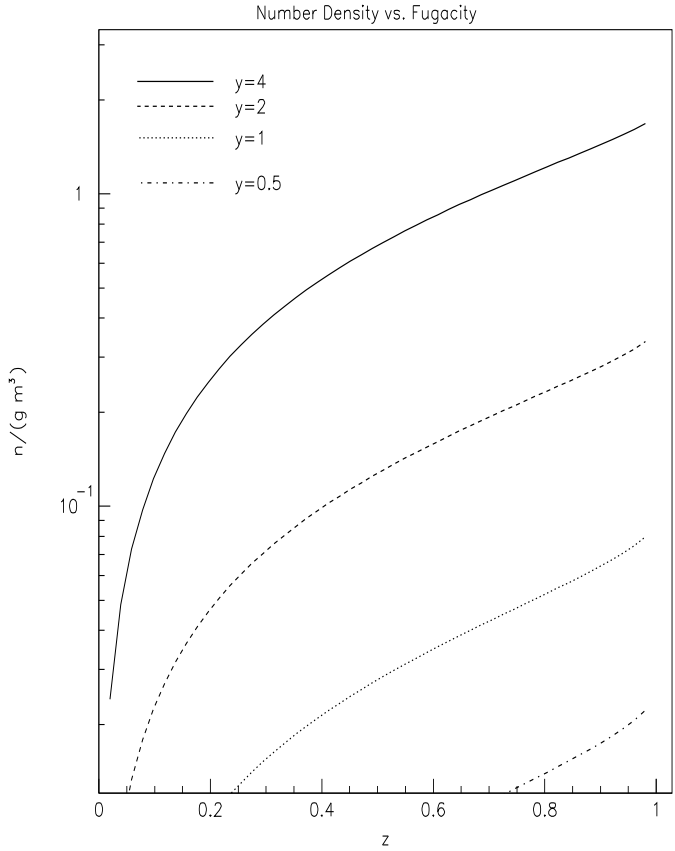,width=4in,height=5in}
\caption{\label{densities} Number density for a given species as a
function of fugacity (see text) for various values of the quotient
$y=\frac{m_\pi}{T}$.}
\end{figure}
\end{widetext}

First we give the number density for particle of species $a$:
\ba \label{density}
n_a= \frac{g_a}{2\pi}\int_0^\infty dp p^2 \frac{1}{e^{-\beta(\mu_a-E)}-1}
\\ \nonumber
+ \frac{g_a}{V} \frac{1}{e^{-\beta(\mu_a-m_a)}-1} \ .
\ea
The partial pressure for species $a$ reads
\ba
P_a= \frac{-g_a T}{2\pi^2} \int_0^\infty dp p^2 \log \left[
1-e^{\beta(\mu_a-E)} \right] \\ \nonumber
- \frac{g_aT}{V} \log \left[  1-e^{\beta \mu_a}\right] 
\ea
and the total pressure is simply given by the sum of the partial
pressures
\be
P=\sum_a P_a \ .
\ee
We will immediately drop the Bose-Einstein condensate term since it
is only relevant at essentially zero temperature, and keep only the
integral over the state continuum in both expressions. Figure 
\ref{densities} shows the number density normalized to make the
species-independent quantity $n/(gm^3)$ as a function of the fugacity.

The internal energy density per species is then 
\be
\epsilon_a=\frac{g_a}{2\pi^2} \int_0^\infty dp p^2 
\frac{E}{e^{-\beta(\mu_a-E)}-1} \ .
\ee
In the absence of a chemical potential it scales as $T^4$. 
Corrections due to interactions (which we consider second 
order) are to be found e. g. in \cite{gerber}. 
We will operate
with the convenient shorthanded adimensional variables $x$, $y$, $z$:
\ba \label{xyz}
x_a= \frac{p^2}{m_a^2} \\ \nonumber
E=m_a\sqrt{1+x} \\ \nonumber
y_a=\beta m_a= \frac{m_a}{T} \\ \nonumber
z_a=e^{\beta(\mu_a-m_a)} \ .
\ea
Defining two simple functions through quadrature:
\ba
l_n(y,z)= \int_0^\infty \frac{x^n dx}{\sqrt{1+x}
\left( z^{-1} e^{y(\sqrt{1+x}-1)}-1\right)}
\\ 
t_n(y,z)= \int_0^\infty \frac{x^n dx}{
\left( z^{-1} e^{y(\sqrt{1+x}-1)}-1\right)}
\ea
which satisfy a recursion relation
\be
t_{n-1}(y,z) =\frac{yz}{2n} \frac{\partial l_n(y,z)}{\partial z}
\ee
the expressions for the partial pressure and number density
become very compact:
\ba
P_a=\frac{g_a m_a^4}{12 \pi^2} l_{3/2}(y_a,z_a) \\ 
n_a=\frac{g_am_a^3}{4\pi^2}t_{1/2}(y_a,z_a) \ .
\ea

Therefore the equation of state for a one component gas is
\be \label{stateeq}
P_a=\frac{m_a}{3} \frac{l_{3/2}(y_a,z_a)}{t_{1/2}(y_a,z_a)} n_a
\ee
and summing over all species we obtain the final equation of state for a 
multicomponent ideal gas in thermodynamic equilibrium
\be \label{eqstate}
P=\sum_a m_a n_a(y_a,z_a) \frac{l_{3/2}(y_a,z_a)}{3 t_{1/2}(y_a,z_a)}
\ee
in which the independent variables can be taken to be $T$, $z_a$,
or $T$, $n_a$ by use of (\ref{density}).
Equation (\ref{eqstate}) is a generalization of the classical
(non-relativistic), ideal mixed gas state equation
$$
P = T \sum_a n_a \ .
$$

\subsection{Kinetic Theory.}
The distribution functions $f_i$ satisfy an Uehling-Uhlenbeck equation. 
This is a Boltzmann-type equation which includes quantum Bose-Einstein
statistics in the final state. Since we consider three particle
species, we have a set of coupled equations:
\ba \label{transport}
\frac{df_\pi}{dt} &=& C[f_\pi,f_\pi] + C[f_\pi,f_K] + C[f_\pi,f_\eta]
 \nonumber \\
\frac{df_K}{dt} &=& C[f_K,f_\pi] \\ \nonumber
\frac{df_\eta}{dt} &=& C[f_\eta,f_\pi]
\
\ea
neglecting $KK$, $K\eta$, $\eta \eta$ as well as inelastic interactions,
with the collision term
\ba
C[f_a,f_b] = \int d\sigma_{ab}d\vec{p}_1 v_{\rm rel} \left[
f_a' f_{1b}'(1+\xi_a f_a)(1+\xi_bf_{1b}) \right. \nonumber \\ \left.
-f_af_{1b} (1+\xi_a f_a')(1+\xi_bf_{1b}') \right] \ .
\ea
Here $v_{\rm rel}$ is the relative velocity between the colliding
particles and in terms of the Lorentz-invariant square of the 
scattering amplitude and phase space, we have
\be
d\sigma_{ab} = \frac{1}{4E(p)E_1v_{\rm rel}} \ar T \ar^2 
d{\rm LIPS}(s;\vec{p} \ ' , \vec{p}_1\ ' )
\ee 
with 
\be
d {\rm LIPS}= \frac{1}{4\pi^2} \frac{p'dE'}{E_1'} d\Omega(p')
\delta(E(p)+E_1-E'-E_1') \ .
\ee
For pion-pion collisions, since the particles in the final state
are undistinguishable to the strong interactions in the isospin
limit, a factor of $\frac{1}{2}$ should multiply the 
integral over $d\sigma_{\pi\pi}$.

The collision term is annihilated by the Bose-Einstein distribution 
function which corresponds to a gas in thermal equilibrium
\be
f_{0a}= \frac{\xi_a^{-1}}{z^{-1} e^{\beta(E(p)-m_a)}-1} \ .
\ee
Small departures from equilibrium are conventionally denoted
by 
\ba \label{chapmanenskog}
f_a=f_{0a}+\delta f_{0a} \\ \nonumber
  =f_{0a}\left( 1+\frac{\chi_{0a}}{T} \right) \ .
\ea
The particle and energy densities can be obtained by integrating over 
momenta:
\ba
n(\vec{x})=\int d\vec{p}f_0 \ , \ \  \rho=\int d\vec{p} f_0 E(p)  \ .
\ea

The contribution to the stress-energy tensor can be shown to be, 
summing  over particle species:
\be
\tau_{ij}= \sum_a \int \frac{d\vec{p}}{E}
p_i p_j \delta f_a \ .
\ee

For the purpose of evaluating the shear viscosity, which appears
at the hydrodynamic level as a tensor of first order in the velocity
gradient, the perturbation is taken to be of the form
$$
\chi_a= g^{ij}_a\tilde{V}_{ij}
$$
with $\tilde{V}$ defined in (\ref{3Dstresstensor}) above
and as a consequence of the contraction with a traceless tensor,
only the traceless part of $g$ is relevant, so we can also take
\be \label{defgdep}
g^a_{ij}=(p_ip_j-\frac{1}{3}\delta_{ij}p^2) g^a(p)
\ee
with $g(p)$ a scalar function, conveniently expanded in a polynomial base
\be \label{gexpand}
g_a(x)=\sum_{s=0}^\infty B_s^{(a)} P^s(x;y,z) \ ,
\ee
the polynomials $P$ being defined in the appendix.

\section{Solution of the Transport Equation}

We next show how the transport equations (\ref{transport}) can be simply 
solved near equilibrium. 
We  describe the method for only one particle species and 
leave the generalization to three or more to the reader. 

\subsection{Advective Term.}

To simplify the left hand side in (\ref{transport}) write it as
\be \label{advective}
\frac{df}{dt}=\frac{\partial f}{\partial t} + \frac{\vec{p}}{E(p)} \cd 
\vec{\nabla} f
\ee
($\vec{v}=\vec{p}/E(p)$)
with an approximately constant perturbation: 
$$
\partial_t f \simeq \partial_t f_0 \ .
$$

In the proper Eulerian frame we know $\vec{V}=0$ but this does not
apply to its derivatives $\partial_i \vec{V}$ which are in general
not null. This forces us to Lorentz transform to an arbitrary, nearby 
frame in this  derivation:
$$
f_0= \frac{\xi^{-1}}{\left[ e^{-\mu/T} 
e^{(E(p)\sqrt{1-V^2}-\vec{p}\cd\vec{V})/T}
-1 \right] }
$$
with $E(p)=\sqrt{p^2+m^2}$ and $\mu \leq m$.
Thus, in (\ref{advective}),
\ba
\frac{\partial f}{\partial t} = \frac{f_0}{T} \frac{1}
{1-e^{-\beta(E(p)-\mu)}} \left( \vec{p}\cd \frac{\partial \vec{V}}
{\partial t} \right. \\ \nonumber \left.
+ \left. \frac{\partial \mu}{\partial P}\right\arrowvert_T
\frac{\partial P}{\partial t} +
\left[  \left. \frac{\partial \mu}{\partial T}\right\arrowvert_P
-\frac{\mu-E(p)}{T} \right] \frac{\partial T}{\partial t}\right) \ .
\ea
This can be reduced \cite{landau} by using the standard thermodynamical
relations, valid near equilibrium to first order in the perturbation,
such as the Maxwell identities, to the form
\ba
\frac{\partial f}{\partial t} = \frac{f_0}{T} \frac{1}
{1-e^{-\beta(E(p)-\mu)}} \left( \vec{p}\cd \frac{\partial \vec{V}}
{\partial t} \right. \\ \nonumber \left.
+\frac{1}{n} \frac{\partial P}{\partial t} + \beta(E(p)-\omega/n)
\frac{\partial T}{\partial t}
\right)
\ea
and in an analogous way

\ba
\vec{p}\cd \vec{\nabla}f = \frac{f_0}{T} \frac{1}
{1-e^{-\beta(E(p)-\mu)}} \left( p_ip_j V_{ij}^{ } \right. \\ \nonumber 
\left. + \frac{1}{n} \vec{p}\cd \vec{\nabla}P+ 
\beta(E(p)-\omega/n)\vec{p}\cd
\vec{\nabla} T \right) \ .
\ea

Next we profit again of the vicinity to equilibrium and employ the
equation of continuity (\ref{continuity}) and state equation 
(\ref{stateeq}) , while we ignore all terms proportional to 
$\vec{\nabla}T$, $\vec{\nabla}P$,
$\vec{\nabla}\cd \vec{V}$ since they influence the calculation of the
volume viscosity or thermal conductivity, but not the shear viscosity
upon comparison with (\ref{3Dstresstensor}) to obtain the final expression
for the advective term:
\be \label{finaladvective}
\frac{d f}{d t} = \frac{f_0}{TE} \frac{1}{1-e^{-\beta(E(p)-\mu)}}
p_ip_j \tilde{V}^{ij} \ .
\ee

\subsection{Collision Term.}
The integrand of any of the collision functionals $C[f_a,f_b]$
contains the product of Bose-Einstein distributions with
enhanced final state phase space:
\be
F= f'_a f'_{1b}(1+\xi_a f_a)(1+\xi_b f_{1b}) - f_a f_{1b}(1+\xi_a
f'_a)(1+\xi_b f'_{1b}) \ .
\ee
For perturbations near local equilibrium, the distribution function 
can be written as in eq. (\ref{chapmanenskog}) and this entails for
the integrand
$$
F=F_0+\delta F \ .
$$
The collision functional evaluated on the Bose-Einstein equilibrium
distribution is naturally zero:
\be
C[f_{0a},f_{0b}]=0
\ee
After some simple algebra we can show
\ba \\ \nonumber
\delta F = \frac{f'_{0a}f'_{01b}f_{0a}f_{01b}}{T\xi_a\xi_b}
e^{\beta (E-\mu_a-\mu_b)} \Delta[\chi (1-e^{-\beta(E-\mu)})] \\  \nonumber
\Delta[X]=X'_a + X'_{1b}-X_a - X_{1b}  
\ea
and it finally follows
\ba \label{finalcollision}
C[f_a,f_b]= \frac{\xi_a\xi_b}{T z_a z_b} \int d \sigma_{ab} d\vec{p}_1
v_{\rm rel} e^{\beta E} \\ \nonumber
f'_{0a} f'_{01b} f_{0a} f_{01b} \Delta[\chi(1-e^\beta(E-\mu))]
\ea

\subsection{Expression for the Viscosity.}
Upon substitution of (\ref{gexpand}) on (\ref{finalcollision}), and 
neglecting $\delta f$ against $f_0$ on the advective term 
(\ref{finaladvective})  the linearized 
transport equation  takes the form
\ba \label{supersistema}
\left(
\begin{array}{rcc}
A^\pi_{11}+ A^K_{11}+A^\eta_{11} & A_{12} & A_{13} \\
A_{21} & A_{22} & 0 \\
A_{31} & 0 & A_{33}
\end{array} \right) \cd \left(
\begin{array}{c}
B_0^\pi \\
B_0^K \\
B_0^\eta
\end{array} \right)
\\ \nonumber
= \left( \begin{array}{c}
C_{\pi} \\
C_K \\
C_{\eta} 
\end{array} \right)
\ea
Once the integrals in $A$'s (this is the largest computation) and $C$'s 
have been performed, the matrix system
can be solved and the $B$ coefficients in (\ref{gexpand}) provide the 
first approximation
to the function $g(p)$. The series converges very fast as shown in
\cite{silvia} and therefore we keep only this order in the present
work. This can then be substituted in the viscosity 
expressions below which can then be easily integrated on a computer.
\be
\eta= -\frac{1}{10 T} \sum_{a}\int \frac{d\vec{p} \ '}{E'}
p_i' p_j' f_{0a}' g^{'ij}_a
\ee
(compare with expression 4.19 in reference \cite{prakash4}).
The angular integrals can be trivially performed, reducing
the expression to
\be
\eta= -\frac{4\pi}{15T} \sum_a \int \frac{dp \ '}{E'}
p'^6f_{0a}'g_a(p') 
\ee
which in terms of dimensionless variables (those of $g$ can be 
read off eq. (\ref{chapmanenskog}) ) gives
\begin{widetext}
\be
\eta=-\frac{2\pi}{15T\xi_a} \sum_a m_a^6  \int_0^\infty dx 
\frac{x^{5/2}g_a(x)}{\sqrt{1+x}\left( z^{-1}_a e^{y_a(\sqrt{1+x}-1)} -1
\right)}
\ee
Employing (\ref{gexpand}) we can write
\be
\eta=-\frac{2\pi}{15 T} \sum_a \sum_{S=0}^{\infty} B_s^{(a)}
\frac{m_a^6}{\xi_a} \int_0^\infty dx W_{5/2}(x_a;y_a,z_a) P_s(z_a;x_a,y_a)
\ee
and using the orthogonality relation (\ref{ortho})
\be \label{finalvisc}
\eta=-\frac{2\pi}{15T} \sum_{a=1}^3 B_0^{(a)} \frac{m_a^6}{\xi_a}
A^0_{5/2(y_a,z_a)}
\ee
\end{widetext}
\section{Meson Scattering.}
With our kinematical choice of variables (see section 
\ref{variousdefs}) we can express the square of the scattering
amplitude for $\pi$-$\pi$ collisions as
\ba \label{pipiamplitude}
 \ar T \ar^2 = \frac{1}{9} (\ar T_0\ar^2+3\ar T_1\ar^2 +5 \ar T_2\ar^2)=
  \frac{1}{9}\left( \frac{32 \pi}{\rho_{\pi\pi}}
\right)^2 \cd  \\ \nonumber
\left(  \sin^2 \delta_{00}(s) +  
27  \left( 1+ \frac{2t}{s\rho_{\pi\pi}^2} \right)^2 
\sin^2 \delta_{11}(s)+ 5\sin^2\delta_{20} \right)
\ea
where the isospin 0,1 or 2 has been averaged and only the lowest partial
waves are kept, since they provide most of the interaction at low
energies.
For pion/kaon collisions we can also average over isospin to obtain:
\ba
 \ar T_{KK} \ar^2 = \frac{1}{6} (2\ar T_{1/2}\ar^2+4\ar T_{3/2}\ar^2)=
\\ \nonumber
 \frac{1}{3}\left(\frac{16\pi}{\rho_{\pi K}}\right)^2 \left(
2\sin^2\delta_{\frac{3}{2} 0} + \sin^2 \delta_{\frac{1}{2} 0}  +
9\sin^2 \delta_{\frac{1}{2} 1}
\left( 1+ \frac{2t}{s\rho_{\pi K}^2}\right)^2
\right) 
\ea
and finally for $\eta$-$\pi$ scattering
\be
\ar T_{\eta \pi}\ar^2 = \left( \frac{16 \pi}{\rho_{\pi \eta}}
\right)^2 \sin^2\delta_{10} \ .
\ee
Notice the partial wave normalization gives $(16)$ for distinguishable
and $(32)$ for identical particles in agreement with \cite{prakash4}.
If only a pion gas is considered, the relevant phase shifts
are the $J=0$, $I=0,2$ and $I=1$, $J=1$ which dominate the cross
section at low energy.
We next study the pion, and then the meson gas, in various approximations
which are
\begin{enumerate}
\item the approximation studied in \cite{prakash3} corresponds to taking
a simple resonance saturation parametrization for the isoscalar and 
isovector phase shifts:
\ba  \label{prakashphases}
\delta_{00}(E)= \frac{\pi}{2}+\arctan \left( 
\frac{E-m_\sigma}{\Gamma_\sigma/2} \right) \\ \nonumber
\delta_{11}(E)=\frac{\pi}{2}+\arctan \left( \frac{E-m_\rho}{\Gamma_\rho/2}
\right)
\ea 
with momentum dependent $\sigma$, width $\Gamma_\sigma=2.06 \ p$
and mass $m_\sigma=5.8 \ m_\pi$, and $\rho$ width
$$
\Gamma_\rho(p)=0.095 p \left( \frac{p/m_\pi}{1+(p/m_\rho)^2}
\right)^2 \ , 
$$
and mass $m_\rho= 5.53 \ m_\pi$ 
whereas the scalar isotensor phase shift is simply parameterized
as a straight line
\be
\delta_{20}=-0.12 p/m_\pi \ .
\ee

\item To connect with the non-relativistic calculation, we will also 
shortly employ a (totally unrealistic at moderate temperatures 
above say a few MeV) pion scattering amplitude based on the low
energy theorem of Weinberg. This is
\be \label{weinberg}
\ar T \ar^2 = \frac{23}{3} \frac{m_\pi^4}{f_\pi^4} \ .
\ee 

\item Next,  through our calculation we profit from the IAM (the Inverse 
Amplitude Method \cite{dobado}) fit to 
the pion scattering phase shifts. Whether in the pion gas or in
the mixed pion, kaon and eta gas, chiral perturbation theory provides
an expansion of the scattering amplitudes at low energy:
\be \label{chiralamp}
t_{IJ} = t^{(2)}_{IJ} + t^{(4)}_{IJ} + ...
\ee
where the order of a term in the expansion counts the powers of
$m_\pi$ or momentum, or equivalently inverse powers of $f_\pi$.
The single channel inverse amplitude method constructs a model
amplitude based on this low energy expansion which, incorporating
exact (as opposed to perturbative order by order) unitarity above
the two particle threshold and below inelastic thresholds, allows
to extend chiral perturbation theory providing satisfactory fits
to the scattering amplitudes up to around 1 GeV.
The one-channel IAM amplitude is
\be
t_{IJ} = \frac{t^{(2)}_{IJ}}{1-\frac{t^{(4)}_{IJ}}{t^{(2)}_{IJ}}}
\ee
and can be understood as the [1,1] Pad\'e approximant corresponding
to the Taylor series (\ref{chiralamp}). The first order
follows from Weinberg's low energy theorem. The second order includes
chiral perturbation theory meson loops and order $p^4$ counterterms.
These have a series of coefficients, usually denoted $l_i$ after the
work of Gasser and Leutwyler \cite{gasser} which can at the moment
only approximately be computed theoretically \cite{conpedro}. These
coefficients are in practice fitted to the pion scattering amplitudes
or other observables. Since this rational approximation to the 
scattering amplitudes presents in general poles (in particular the
$\sigma$ and $\rho$ mesons are clearly visible) it has been applied
recently to the behaviour of resonances in a thermal pion bath
\cite{iamsu2}. Here we employ the fit of that paper to the scattering
amplitudes at zero temperature, that is, we ignore the possible effect
of the thermal bath on the parameters entering the fits ($l$'s, $f_\pi$
and $m_\pi$ which are adjusted to their physical value) as a higher
order effect. The $l$'s employed in our $SU(2)$ fits to the pion 
scattering amplitudes are given in table \ref{tablasu2}.

\begin{table}[h]
\caption{\label{tablasu2} Values of the $SU(2)$ chiral perturbation theory
parameters employed in the IAM fit to the pion scattering amplitudes 
(input to this calculation of the viscosity). 
}
\begin{tabular}{|c|c|} \hline
$\ov{l}_1$ & -0.27 \\
$\ov{l}_2$ & 5.56  \\
$\ov{l}_3$ & 3.4  \\
$\ov{l}_4$ & 4.3 \\
\hline
\end{tabular}
\end{table}

\item  Finally, the full calculation including kaons and etas requires
a parameterization of the elastic $SU(3)$ phase shifts. This is 
again provided by chiral perturbation theory now with the strange
quark incorporated \cite{gasser2} and extended to higher energies
via a unitarization method, now the coupled-channel IAM. We use
the phase shifts obtained in \cite{iamsu3} with the $L$ chiral 
perturbation theory parameters shown in table \ref{tablasu3}.

\begin{table}[h]
\caption{\label{tablasu3} Values of the $SU(3)$ chiral perturbation theory
parameters employed in the IAM fit to the meson scattering amplitudes
(input to this calculation of the viscosity).
}
\begin{tabular}{|cc|cc|} \hline
$\ov{L}_1$ & 0.59 & $\ov{L}_5$ & 1.8\\
$\ov{L}_2$ & 1.18 & $\ov{L}_6$ & 0.006\\
$\ov{L}_3$ & -2.93& $\ov{L}_7$ & -0.12\\
$\ov{L}_4$ &  0.2 & $\ov{L}_8$ & 0.78 \\ \hline
\end{tabular}
\end{table}
The use of the IAM is of course not necessary to obtain a
good description of the transport coefficients, but it is
a very convenient parameterization of the experimental data,
providing outstanding fits to the complete set of meson scattering 
channels described in
\cite{iamsu3}.
\end{enumerate}
We finally quote the meson masses employed in this paper: $m_\pi=139.57$ 
MeV, $m_K= 493.677$ MeV, $m_\eta=547.30$ MeV.

\section{Numerical Evaluation}
A number of integrals need to be numerically evaluated to complete
the calculation. Those of the type (\ref{ortho}) needed to compute the 
coefficients $C$ in equation (\ref{supersistema}) as well as the 
viscosity in (\ref{finalvisc}) are 
simple
one-dimensional integrals. The Bose-Einstein factors ensure very
good convergence at high momentum. They are evaluated in a one-dimensional
grid with 2000 points by an open trapezoidal rule in essentially no
computer time. 
More complicated are the integrals stemming from the collision term
of the Uehling-Uhlenbeck equation needed for the $A_{ab}$ coefficients in
(\ref{supersistema}). 
Since we have already chosen the
system of reference where all magnitudes are expressed in the comoving
fluid frame, we have no freedom to study the collision in the center of
mass, so we employ arbitrary momenta in these integrals.
The nominal variables are
$\vec{p}$, $\vec{p}_1$, and $\Omega(\vec{p} \ ')$, while $\vec{p}_1 \ '$
is fixed by momentum conservation and $\ar \vec{p} \ '\ar$ by 
energy conservation.
The later has a somewhat messy expression, in terms of the angle
$\beta$ between $\vec{p} \ '$ and $\vec{P}$:
\begin{widetext}
\ba \label{dossols} \nonumber
p' = \frac{\left(1-\frac{P^2-m_i^2+m_j^2}{E^2} \right)P \cos \beta
\pm \sqrt{\left( 1-\frac{P^2-m_i^2+m_j^2}{E^2} \right)^2P^2 \cos^2\beta
-4 \left( 1-\frac{P^2 \cos^2 \beta}{E^2}  \right)\left( 
m_i^2-\frac{(E^2-P^2+m_i^2-m_j^2)^2}{4E^2} \right)  }}
{2\left( 1- \frac{p^2 \cos^2 \beta}{E^2} \right)} \\ 
\cos \beta = \frac{1}{P} \left( p_1 \sin \theta(\vec{p}_1) 
\sin \theta(\vec{p}\ ') \cos \phi(\vec{p}\ ') 
+\cos \theta(\vec{p}\ ')(p+p_1\cos \theta(\vec{p}_1) )
\right)
\ea
\end{widetext}
Two solutions of this quadratic energy equation
are possible and need to be summed over in the integrand for certain
kinematical configurations.
Of the eight remaining integrals, three are trivial. 
Invariance of the collision under three-dimensional rotations allow us
to perform the angular integrals associated to, say, $\vec{p}$ and refer 
all angles to the $p$-axis. Then
there is still an axial symmetry of the other three particles around this 
axis which allows to perform the azimuthal integral over $\phi(p_1)$. The 
remaining five variables are then $\ar\vec{p}\ar$, $\ar\vec{p}_1\ar$, 
$\theta(p_1)$, $\Omega(\vec{p}\ ')$. The resulting integral is 
performed numerically with the Vegas \cite{lepage} random-point algorithm.
Again the Bose-Einstein factors concentrate the integrand in a compact
set and convergence is very fast. For precision around 1/1000 and 
$\chi^2\simeq 1$ it is enough to start with 2000 points and double that
number 5 or 6 times, while evaluating the integral 10 times for each
fixed number of points. 

\section{Pion Gas Viscosity}
\begin{figure}[h]
\psfig{figure=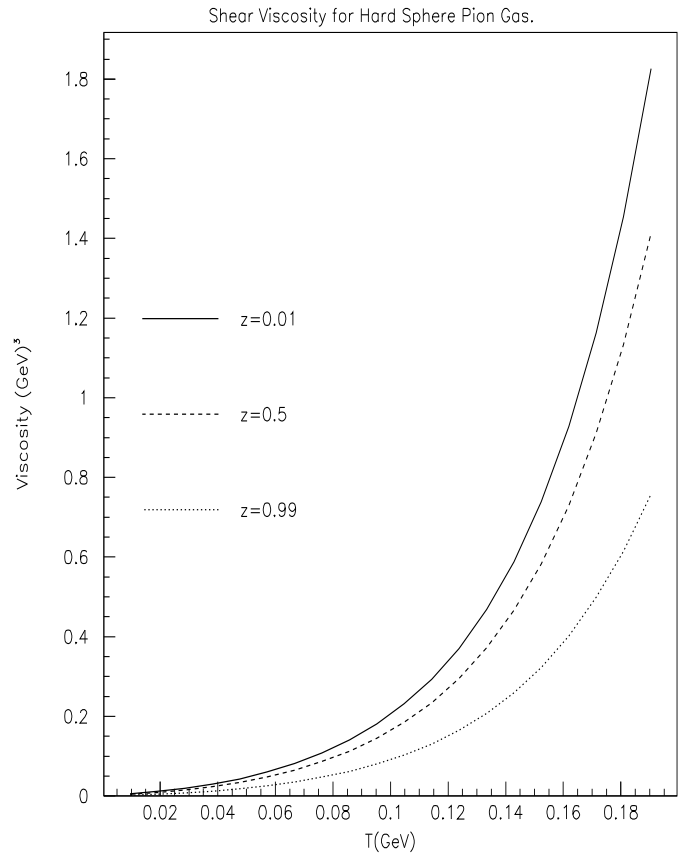,width=4in,height=5in}
\caption{\label{pionhard} Shear viscosity of the pion gas
with a constant scattering amplitude (from Weinberg's theorem).
Since the interaction does not grow with the pion momentum, the
viscosity is unacceptably large even for somewhat low temperatures.
But this is used to check the low temperature limit.
}
\end{figure}

We first evaluate the viscosity with a constant amplitude 
(\ref{weinberg}). The result is plotted in figure \ref{pionhard}.
Since the cross section is independent of energy, an increase in 
temperature simply populates states with faster pions which transfer
momentum more efficiently. Thus, the viscosity grows out of
control in an unrealistic manner. But this calculation is useful
to check the non-relativistic limit reported in \cite{silvia}, which
used precisely this interaction. The behaviour at low temperature
(and this is common to all our calculations) is governed by a non-
analytic $\eta \propto \sqrt{T}$ behaviour. To see it, simply 
remember that for a hard-sphere classical gas, the mean free path is 
inversely proportional to the cross section and density;
$$
\lambda = \frac{1}{\sqrt{2}n\sigma}
$$
and calculating the momentum transfered by random flight of the
gas molecules one can obtain
$$
\eta=\frac{1}{3} n m \sqrt{\ov{v^2}} \lambda
$$
in terms of the rms. velocity. This shows that the viscosity
is inversely proportional to the cross section and upon employing
the equal partitioning of energy $m\ov{v^2}/2 \simeq 3kT/2$ we see
that the viscosity is also proportional to the root of the temperature,
which provides a convenient check.

\begin{figure}[h]
\psfig{figure=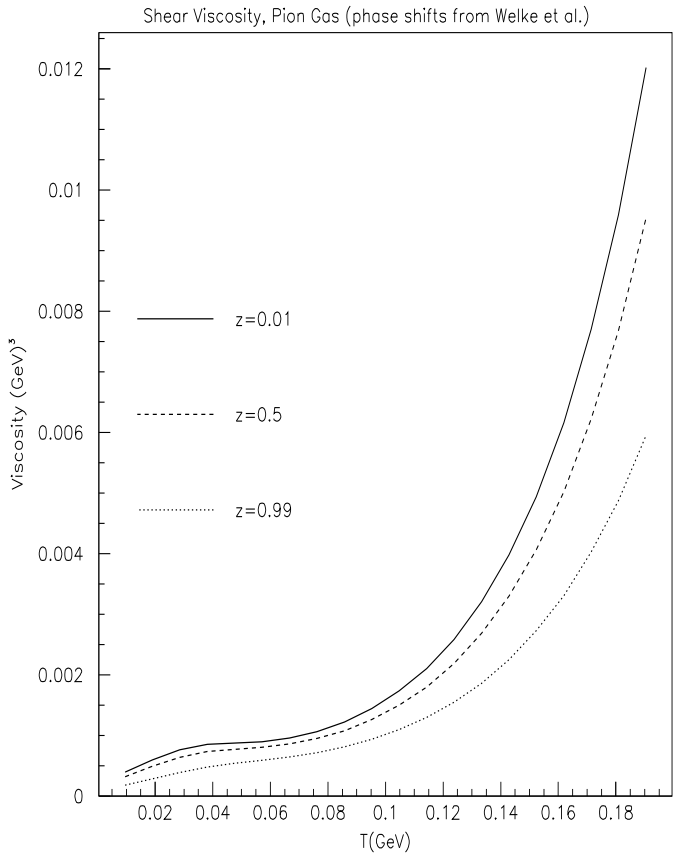,width=4in,height=5in}
\caption{\label{pionprakash} 
Shear viscosity of the pion gas from the simple analytical phase shifts
(\ref{prakashphases}) from Welke et al.}
\end{figure}

\begin{figure}[h]
\psfig{figure=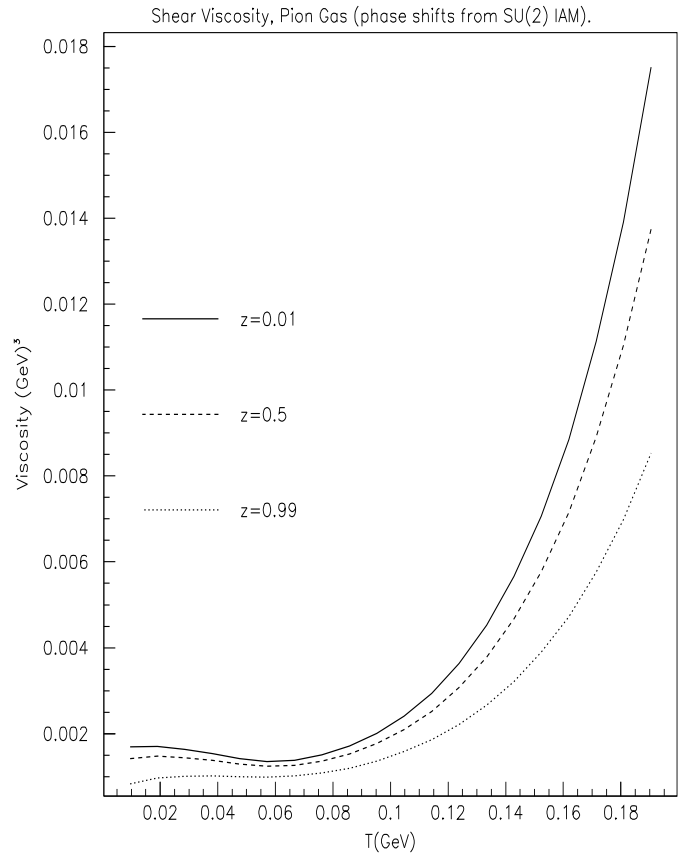,width=4in,height=5in}
\caption{\label{pionsu2}
Shear viscosity of the pion gas from the Inverse Amplitude Method
and $SU(2)$ chiral perturbation theory.}
\end{figure}

\begin{figure}[h]
\psfig{figure=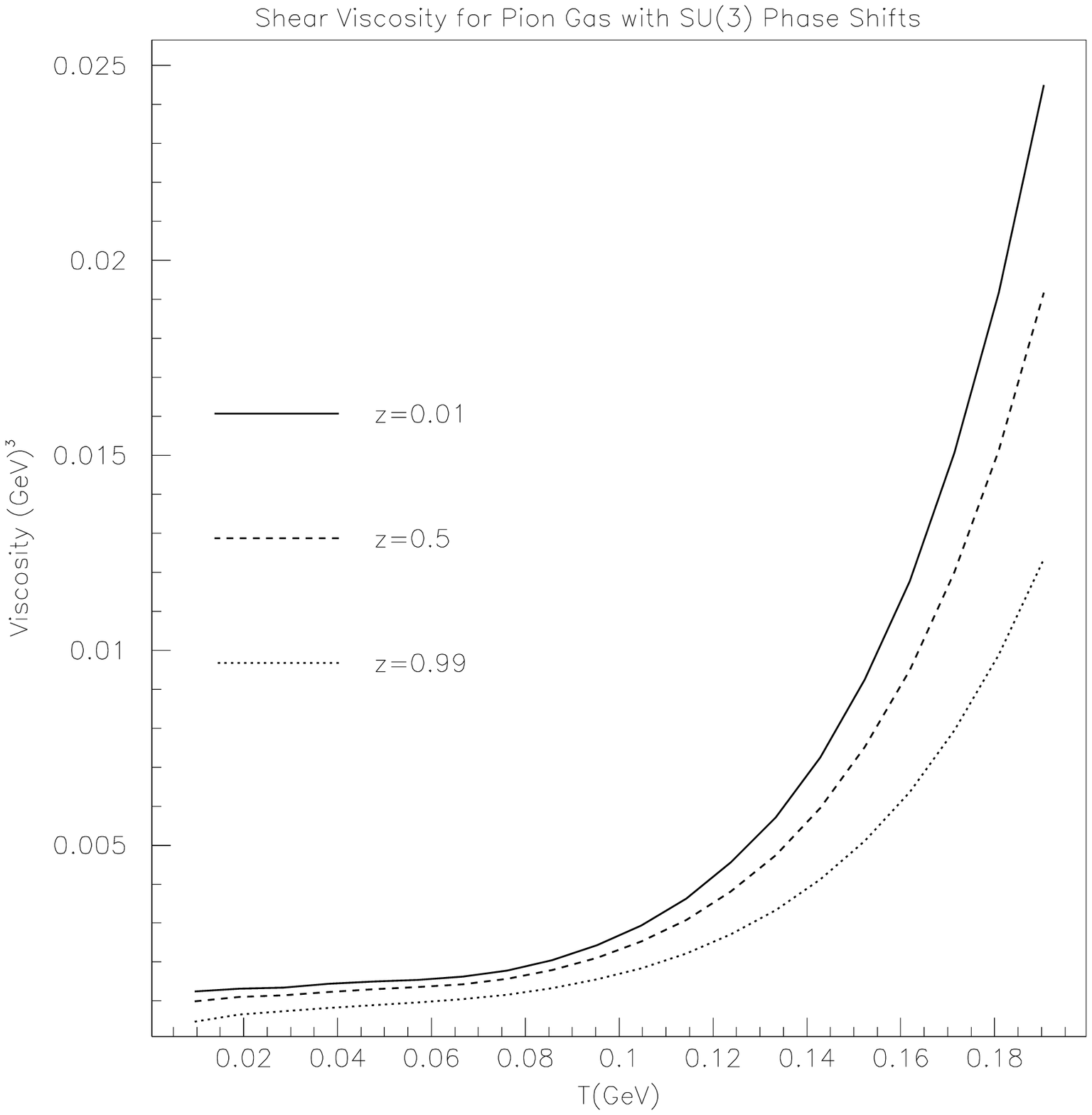,width=4in,height=5in}
\caption{\label{pionsu3} Shear viscosity of the pion gas from the
$SU(3)$ Inverse Amplitude Method phase shifts.}
\end{figure}

Next we turn to some more realistic pion interactions. These are provided
by the simple analytical fit to the pion phase shifts from 
(\ref{prakashphases}) and by the $SU(2)$ or $SU(3)$ inverse amplitude 
method. 
The results are quite consistent and plotted in figures 
\ref{pionprakash}, \ref{pionsu2} and \ref{pionsu3} respectively. The 
difference between them  gives us an idea of the sensitivity of the 
viscosity to the employed phase shifts, since all sets of phase shifts 
are reasonable. To what precision these scattering phase shifts are known
is an ongoing debate \cite{warriors} and if a future determination pinned 
them down
with greater accuracy a much better prediction for the viscosity could
be made, since the parameterization used for the phase shifts seems
to be one of the largest uncertainty sources in the present 
(already realistic) computation.

\section{Full Pion, Kaon, Eta Gas Viscosity.}

\begin{figure}[h]
\psfig{figure=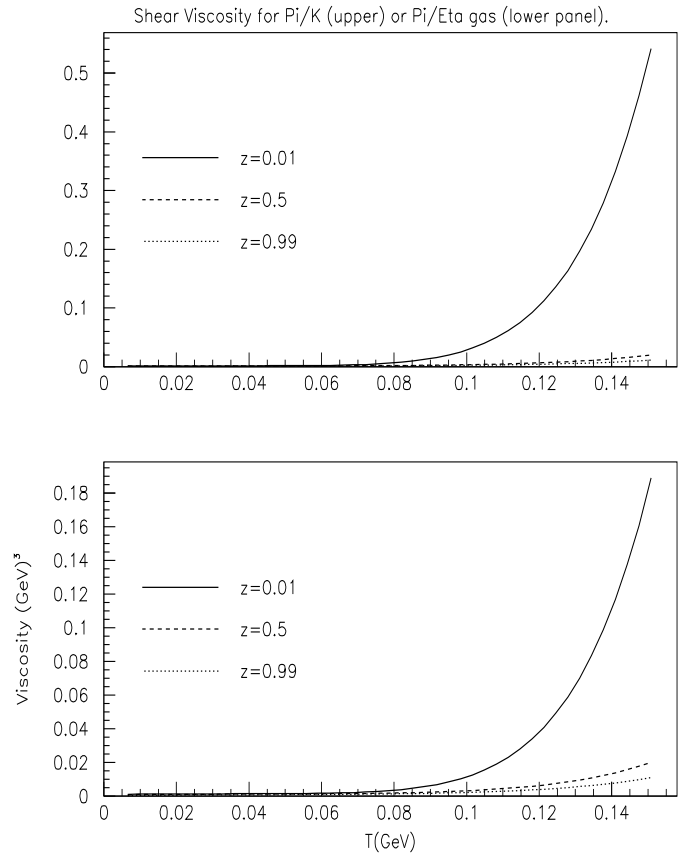,width=4in,height=5in}
\caption{\label{ksoretas}  Separate effect of adding $K$ or $\eta$
mesons to the pion gas, with their chemical potential vanishing. The
kaons are by far more important (partly due to their multiplicity).}
\end{figure}

\begin{figure}[h]
\psfig{figure=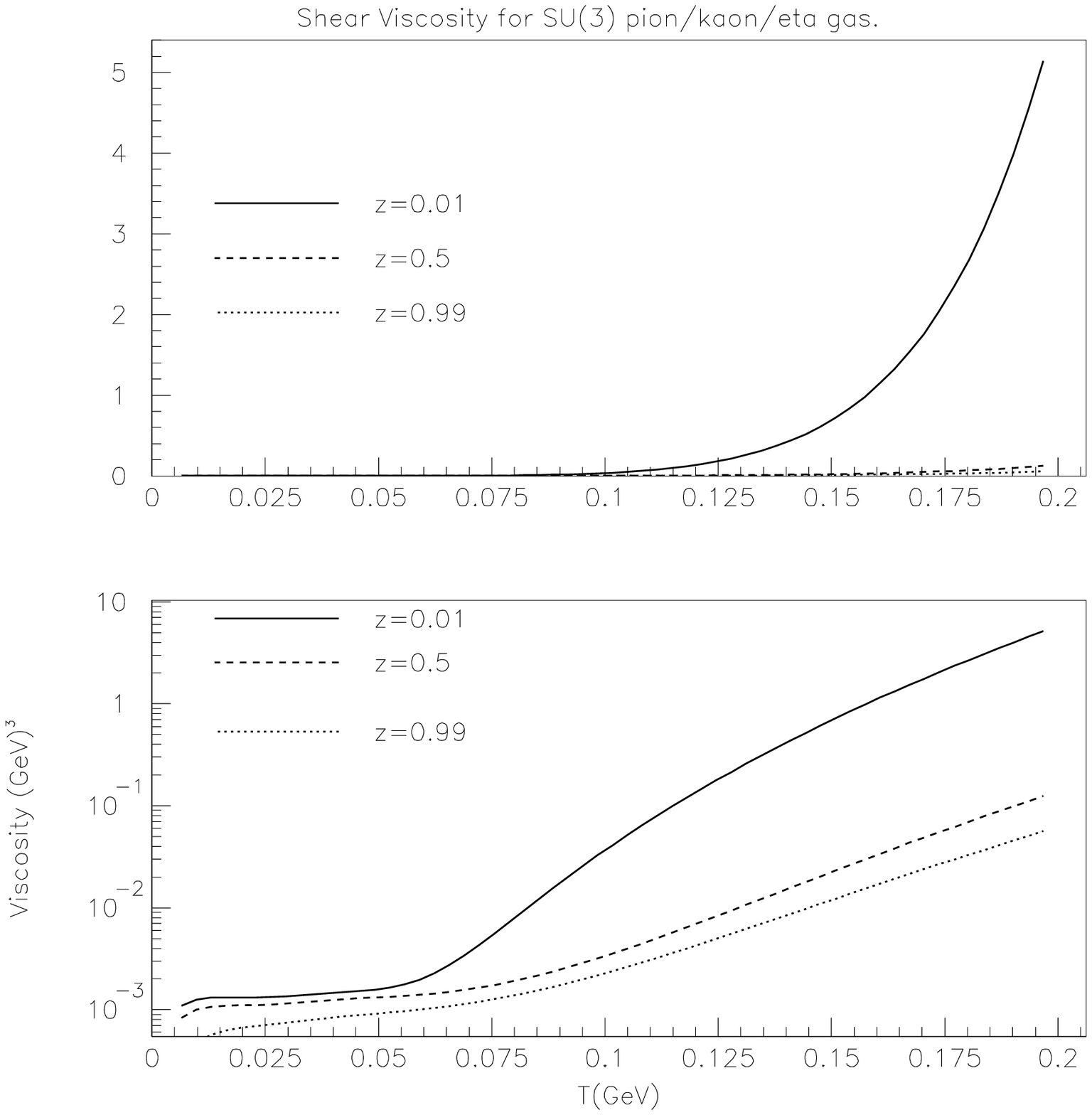,width=4in,height=5in}
\caption{\label{ksandetas} The shear viscosity for the full meson
($\pi$, K, $\eta$) gas in linear (upper) and semilogarithmic (lower)
panels. At very low $z$ and $T$ the reader can appreciate how the
non-relativistic behaviour ($\eta \propto \sqrt{T}$) is recovered.
The chemical potential for both $K$ and $\eta$ has been set to zero.}
\end{figure}

Finally we turn to a gas including kaon and eta mesons. We first
introduce either type of particle separately and plot it in fig.
\ref{ksoretas}. The effect of kaons is much more dramatic than the
effect of eta mesons. By comparing with fig. \ref{pionsu3} which was
calculated with the same pion phase shifts we can see that already at 
100 MeV the addition of the kaons alone give a viscosity way bigger than
present in the pion gas. The sensitivity of this calculation on the pion
fugacity (density) is also very large.
The combined effect of adding both kaons and etas to the gas is
finally plotted in fig. \ref{ksandetas}. 
Of course, in a relativistic 
heavy ion collision we expect the chemical freeze-out of heavier mesons
to occur before and therefore we need to introduce chemical potentials
for all species, those corresponding to the heavier mesons being larger
than for lighter mesons.

\begin{figure}[h]
\psfig{figure=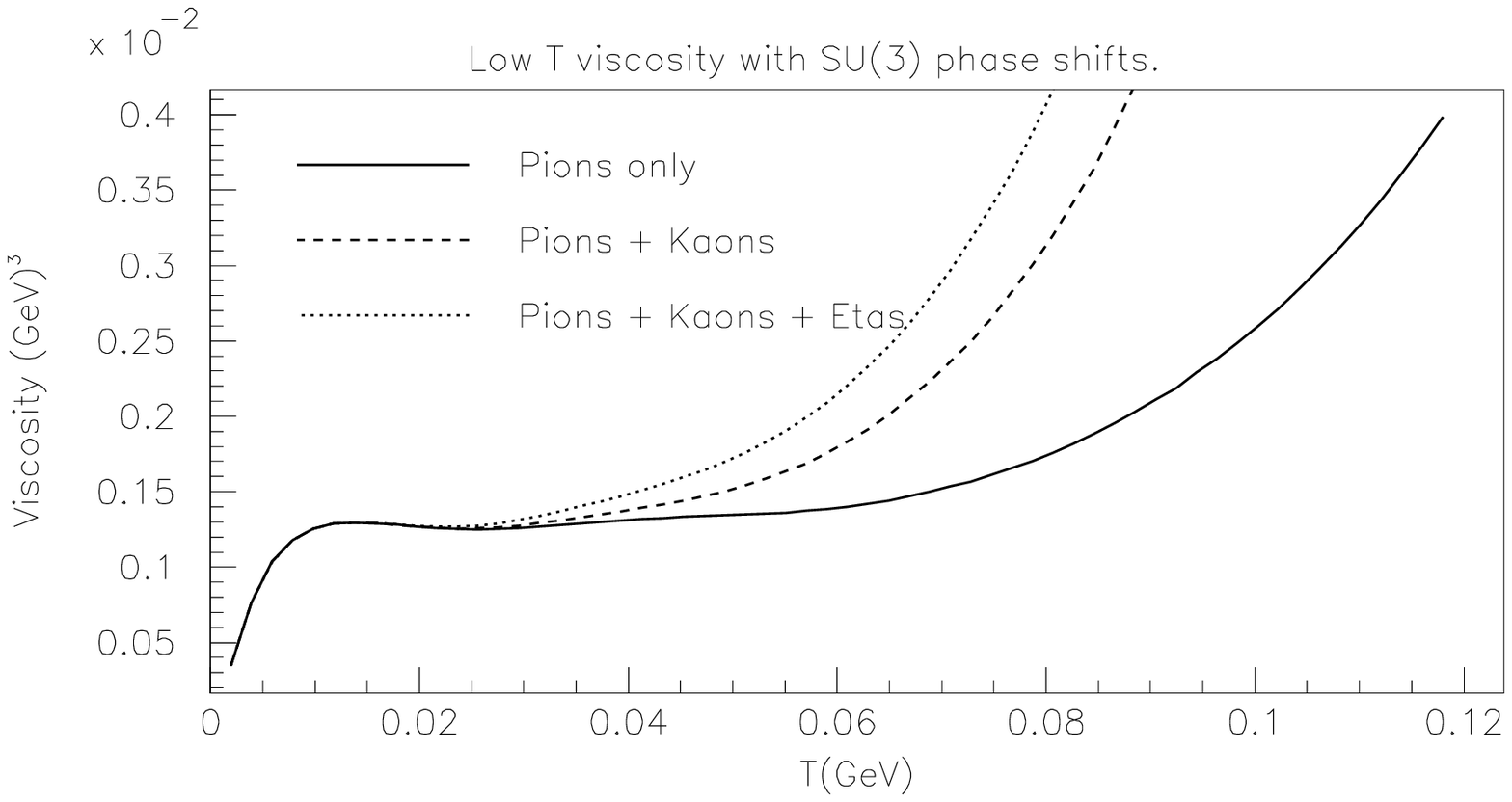,width=4in,height=5in}
\caption{\label{lowTplot} The shear viscosity for the full meson
($\pi$, K, $\eta$) gas at lower temperature.
The chemical potentials are $\mu_\pi=100$ MeV, $\mu_K = 250$ MeV, 
$\mu_\eta= 300$ MeV. The low energy plateau is reduced by adding 
more species, as kaons and etas behave less as Goldstone bosons than
the pion, so their low energy interactions are already larger and don't 
see such a large relative increase as the pion's with collision energy, 
and the viscosity increases more rapidly with the 
temperature as the number of species is increased.}
\end{figure}

In natural units the viscosity has dimensions of an energy cubed, and 
indeed at moderate to high temperatures, the viscosity approximately 
follows a power law with an exponent near (and slightly above) 3, which is
suggesting the viscosity is dominated by  the highest energy pions where 
the mass and chemical potential scales are less important than the 
momentum scale. 
The low temperature behaviour of the viscosity is plotted in figure
\ref{lowTplot} where we observe how at low temperature the effect
of adding the more massive particles is to decrease the plateau
in which the viscosity is approximately independent of the temperature.

The resulting viscosities for various temperatures and chemical potentials 
are tabulated in table \ref{viscosityvalues} and convey our final results.

\begin{table}[h]
\caption{\label{viscosityvalues}
Shear viscosity at various values of chemical potentials and temperature.
Units are MeV ($T$ and $\mu_a$) and $(100 \ {\rm MeV})^3$ ($\eta_s$).
The error due to the Montecarlo evaluation of the 5-D integral is (5) on 
the last significant digit of the viscosity. This calculation employs 
the Inverse Amplitude Method phase shifts within $SU(3)$.}
\begin{ruledtabular}
\begin{tabular}{cccc|cc}
 $T$  &  $\mu_\pi$  &  $\mu_K$  &  $\mu_\eta$  &  $\eta_s$   \\ \hline
100 &   0 &   0 &   0 & 3.37 \\
100 &   0 & 100 & 150 & 5.24 \\
100 &   0 & 200 & 250 & 9.96 \\
100 & 100 & 100 & 150 & 3.04 \\
100 & 100 & 200 & 250 & 4.56 \\
100 & 100 & 400 & 450 & 19.0 \\
125 &   0 &   0 &   0 & 7.13 \\
125 &   0 & 100 & 150 & 11.4 \\
125 &   0 & 200 & 250 & 20.1 \\
125 & 100 & 100 & 150 & 6.31 \\
125 & 100 & 200 & 250 & 9.69 \\
125 & 100 & 400 & 450 & 31.6 \\
150 &   0 &   0 &   0 & 15.4 \\
150 &   0 & 100 & 150 & 24.2 \\
150 &   0 & 200 & 250 & 39.6 \\
150 & 100 & 100 & 150 & 13.4 \\
150 & 100 & 200 & 250 & 20.2 \\
150 & 100 & 400 & 450 & 53.8 \\
\end{tabular}
\end{ruledtabular}
\end{table}

\section{Comparison with Other Approaches and Discussion}
The viscosity of a pion gas has already been treated by D. Davesne 
in \cite{davesne} as a quantum relativistic system. In this pioneering
work to solve the relativistic Uehling-Uhlenbeck equation the pion 
interaction was modelled following ref. \cite{prakash3}, which corresponds 
to the phase shifts (\ref{prakashphases}) above. If we consider only a 
pion gas and employ the same set of phase shifts then we 
can  approximately reproduce this results.
Furthermore, we document the sensitivity of
the viscosity to the parameterization of the phase shifts, which was
not treated in this reference. We have also streamlined the numerical
solution of the transport equation, by employing a new family of 
orthogonal polynomials which allows to systematically extract better 
approximations if so wished, and performing a  Montecarlo evaluation
of the collision multidimensional integral, whereas the more analytical 
treatment in \cite{davesne} is also somewhat more obscure.
It is also interesting that a relaxation time  estimation of the Boltzmann
equation (with no quantum corrections)
permits to approximate the shear viscosity of a pion/kaon gas 
in \cite{prakash4,prakash1}. 
The order of magnitude
and qualitative behaviour as a function of temperature are correct. We 
obtain somewhat larger results which are not unexpected as quantum 
corrections in a Bose gas may tend to decrease the cross section for 
scattering to initially unpopulated states, increasing the viscosity, and
furthermore some difference is expected due to our use of the $SU(3)$ 
phase shifts.

To summarize, we have presented a systematic calculation of the shear
viscosity in meson matter at moderate temperatures. 
We have found the viscosity to behave as expected from a non-relativistic 
gas point of view at very low temperatures, to stabilize and even decrease
(depending on the chemical potential) at small temperatures because
of the larger cross section at increasing energies (the decrease would be 
a typical effect of a Goldstone boson gas) and to follow a positive
power law at moderate to high temperatures.  At high $T$ (above 150 
MeV) our
approach should be less reliable because we are employing scattering phase 
shifts parameterized up to momenta of 1 GeV and with sizeable 
temperatures, states with higher momentum start being populated. 
Eventually one reaches the phase transition temperature, and any
results obtained from within the chirally broken phase (as built-in in
our use of meson fields and chiral perturbation theory) is simply not
appropriate.
Extensions of this 
work to include nucleons or to evaluate other interesting transport 
coefficients are now straight-forward. 

\newpage
\acknowledgments

The authors thank J. R. Pel\'aez and A. G\'omez Nicola for providing us
with their SU(3) phase shifts and useful discussions, S. Santalla and F. 
J. Fern\'andez for extensive checks and  assistance, and D. Davesne for
some interesting comments.
This work was supported by grants FPA 2000-0956, BFM 2002-01003 (Spain).

\newpage
\appendix
\section{Linearized Transport Equation Coefficients}
In this appendix we provide the matrix elements necessary for
the first order Chapman-Enskog solution of the transport 
equation.
The right-hand side of the system (\ref{supersistema}) is
\be
\left(
\begin{array}{c}
C_{\pi} \\
C_{K} \\
C_{\eta}
\end{array}  \right) = \left(
\begin{array}{c}
\frac{4\pi m_\pi^6}{3\xi_\pi} l_{5/2}(y_\pi,z_\pi) \\
\frac{4\pi m_K^6}{3\xi_K} l_{5/2}(y_K,z_K) \\
\frac{4\pi m_\eta^6}{3\xi_\eta} l_{5/2}(y_\eta,z_\eta) 
\end{array} \right) \ . 
\ee
The left-hand side matrix elements can be given by the
formula 
\ba \\ \nonumber
A_{ab}=\frac{\xi_a\xi_b}{z_a z_b} \int d\sigma_{ab} v_{\rm rel}
d\vec{p}d\vec{p}_1 e^{\beta(E-m_a-m_b)}f_{0a}f_{0b1}f_{0a}'f_{0b1}' 
\\ \nonumber
\left(1-z_a e^{\beta(E(p)-m_a)}\right) (\delta_{ik}\delta_{jl})-
\frac{1}{3} \delta_{ij}\delta_{kl}) p_i p_j ({\bf A}_{ab})_{kl}
\ea
in terms of a tensor $({\bf A}_{ab})_{kl}$
which takes the values
\ba  \nonumber
({\bf A^\pi}_{11})_{kl}=p^{'k} p^{'l}(1-e^{-\beta(E'-\mu_\pi)})
-p^k p^l(1-e^{-\beta(E(p)-\mu_\pi)}) 
\\ \nonumber
+p_1^{'k} p_1^{'l}(1-e^{-\beta(E_1'-\mu_\pi)})
-p_1^k p_1^l(1-e^{-\beta(E_1-\mu_\pi)})
\\ \nonumber
({\bf A^K}_{11})_{kl}=({\bf A^\eta}_{11})_{kl}= \\ \nonumber
p^{'k} p^{'l}(1-e^{-\beta(E'-\mu_\pi)})
-p^k p^l(1-e^{-\beta(E(p)-\mu_\pi)})
\\ \nonumber
({\bf A}_{12})_{kl}=p_1^{'k} p_1^{'l}(1-e^{-\beta(E_1'-\mu_K)})
-p_1^k p_1^l(1-e^{-\beta(E_1-\mu_K)})
\\ \nonumber
({\bf A}_{13})_{kl}=p_1^{'k} p_1^{'l}(1-e^{-\beta(E_1'-\mu_\eta)})
-p_1^k p_1^l(1-e^{-\beta(E_1-\mu_\eta)})
\\ \nonumber
({\bf A}_{22})_{kl}= p^{'k} p^{'l}(1-e^{-\beta(E'-\mu_K)})
-p^k p^l(1-e^{-\beta(E(p)-\mu_K)})
\\ \nonumber
({\bf A}_{21})_{kl}=({\bf A}_{31})_{kl}= \\ \nonumber
p_1^{'k} p_1^{'l}(1-e^{-\beta(E_1'-\mu_\pi)})
-p_1^k p_1^l(1-e^{-\beta(E_1-\mu_\pi)})
\\ \nonumber
({\bf A}_{33})_{kl}= p^{'k} p^{'l}(1-e^{-\beta(E'-\mu_\eta)})
-p^k p^l(1-e^{-\beta(E(p)-\mu_\eta)})
\ea

\section{Orthogonal polynomials.}
In solving the Uehling-Uhlenbeck equation with relativistic
kinematics and quantum statistics, one needs to integrate over
the measure
\ba
d\mu_r(x;y,z)= w_r(x;y,z) dx = \\ \nonumber
\frac{x^rdx}{\sqrt{1+x}\left( z^{-1} e^{y(\sqrt{1+x}-1)} -1
\right)}
\ea
With the variable $x$ and parameters $y,z$ defined in (\ref{xyz}) above, 
with
ranges  $z\in(0,1)$, $y\in(0,\infty)$, $x\in (0,\infty)$.
The index $r\geq 1$ takes in typical applications a half-integer 
value due to relativistic kinematics.
It can be easily seen that $d\mu^r$ is a valid measure, positive 
definite,  with  bound integrals
$$
\mu_n= \int_0^\infty dx W_r(x;y,z)x^n < \infty
$$
for $n$ a positive integer. As a consequence we can define
a family of orthogonal polynomials analogous to the Sonine polynomials,
but more appropriate for a relativistic Bose-Einstein gas,
which can conveniently be chosen monic (coefficient of highest
dimension term equals one), denoted
$
P^s(x;,y,z)
$
and with an orthogonalization
\be \label{ortho}
\int_0^\infty dx W_r(x;y,z) P^s_r(x;,y,z) P^t_r(x;,y,z) = \delta^{st}
A^s_r(y,z) \ .
\ee
Since the polynomials are considered monic, $A$ is not unity. 
In the calculation presented we have considered a meson gas
slightly out of equilibrium, where a  good approximation
is achieved by keeping only the first polynomial in the expansion
of $g(p)$ defined in (\ref{defgdep}) above. This has been verified
in \cite{silvia} in the non-relativistic limit. Therefore in
this calculation we only need to evaluate the $A_{00}(y,z)$ function.
\end{document}